\documentclass[preprint]{aastex}
\usepackage{amssymb,amsmath}

\begin{document}

\title{Multi-Wavelength Photometry of the T Tauri Binary V582 Mon (KH 15D): A New Epoch of Occultations}

\author{Diana Windemuth and William Herbst}
\affil{Astronomy Department, Wesleyan University, Middletown, CT
06459}
\email{dwindemuth@wesleyan.edu}

\begin{abstract}

We present multi-wavelength (\emph{VRIJHK}) observations of KH 15D obtained in 2012/13, as well as a master table of standard photometry spanning the years 1967 to 2013. The system is a close, eccentric T Tauri binary embedded in an inclined precessing circumbinary (CB) ring. The most recent data show the continued rise of star B with respect to the trailing edge of the occulting horizon as the system's maximum brightness steadily increases. The wealth of data in time and wavelength domains allows us to track the long-term CCD color evolution of KH 15D. We find that the $V-I$ behavior is consistent with direct and scattered light from the composite color of two stars with slightly different temperatures. There is no evidence for any reddening or bluing associated with extinction or scattering by ISM-sized dust grains. Furthermore, we probe the system's faint phase behavior at near-infrared wavelengths in order to investigate extinction properties of the ring and signatures of a possible shepherding planet sometimes invoked to confine the CB ring at $\sim 5$ AU. The wavelength independence of eclipse depth at second contact is consistent with the ring material being fully opaque to 2.2 $\mu$m. The color-magnitude diagrams demonstrate excess flux in $J$ and $H$ at low light levels, which may be due to the presence of a hot, young Jupiter-mass planet.

\end{abstract}
 
 
\section{Introduction}
\label{intro}
V582 Mon is a close, eccentric ($a=0.13$ AU, $e=0.6$, P = 48.37 d) low-mass pre-main sequence (PMS) binary system undergoing regular occultations on the orbital period by a circumbinary ring. The object, also known as KH 15D, is a member of NGC 2264, a young ($\sim$ 3 Myr) open cluster northwest of the Cone Nebula ($d \sim$ 760 parsecs). Extensive monitoring over the years has revealed many intriguing properties of the system, including a gradual change in eclipse depth, duration, and maximum light, a disappearance of brightness reversals during mid-eclipse, and a significant change from 0\% polarization out of eclipse to 2\% during eclipse  \citep{kh1998,hamilton2005,hamilton2003,agol2004}. Photographic plates from 1913--1951 show that the system exhibited no eclipses with modern characteristics during that era \citep{winn2003}, while data from 1967--1982 demonstrate shallower eclipse depths and brighter maxima at a 180$^{\circ}$ phase shift compared to modern data \citep{johnsonwinn2004}. 

To account for the system's observed properties, investigators interpret KH 15D as a pair of T Tauri stars (TTS) embedded in a vertically thin ring of solids \citep{murrayclay2004,winn2004,herbst2010,hamilton2012}. The ring is inclined to the binary orbit, warped, and radially thin, which allows it to precess as a rigid body on a timescale of $\sim$ 2000 years. The precession of nodes manifests itself as a sky-projected sharp-edged ``screen" advancing across the binary orbit. This induces periodic occultations of the stellar objects, which are designated as star A (K6/K7, \citet{hamilton2001}) and star B (K1, \citet{capelo2012}). The position of the screen with respect to the binary orbit modulates the duration and amount of direct light from the stars. Between 1995--2009, the dominant source of light near maximum was star A. Despite the complete obscuration of both stellar components between 2009--2011, variability continued on the binary period -- albeit with reduced amplitude -- due to scattered light from both stars. The system then suddenly rebrightened in late 2011, revealing the trailing edge of the screen and ushering in a new epoch of occultations in which the slightly earlier-type star B now regularly rises at apastron from behind the obscuring horizon. For visual representation of the ``advancing screen" and diagrams of the system at different configurations, see Figs. 5 and 9 of \citet{winn2006}, as well as Fig. 5 of \citet{capelo2012}.

The emergence of star B is a confirmation of the precessing ring model \citep{murrayclay2004,winn2004,winn2006}. The existence of such an ordered structure in a 3 Myr system raises two major questions, which closely relate to current issues in planet formation: 1) What is the composition of the ring? and 2) How is the ring confined? In particular, while tidal interactions between disk and binary confine the inner edge at $\sim$ 1 AU, the truncation mechanism at outer radii ($\sim$ 5--10 AU; \citet{murrayclay2004,lodato2013}) of the CB ring remains unknown. Confinement at the outer region may be achieved by a third body, e.g., a giant planet, in analogy to shepherd moons which maintain planetary rings. However, other mechanisms may explain ring formation (e.g. \citet{lyra2013}).

To address these questions, we obtained multi-wavelength optical and near-infrared (O/NIR) observations of KH 15D during the 2012/13 observing season and append the results to a table of the most complete set of standard photometric data to date. In particular, we investigate the system at intermediate and faint phases, when the stellar photosphere has completely set behind the occulting horizon. We report on the system's periodic (due to reflex motion of the star) and secular (due to precession/architecture of the ring) photometric variabilities as star B now regularly rises above and sets behind the occulting horizon. We supplement our analysis with optical and near-infrared light curves from previous epochs and discuss our results in the context of the long-term color evolution of the system. 
 
\section{Observations}
\label{obs}
Simultaneous O/NIR observations were acquired in queue mode with A Novel Duel Imaging CAMera (ANDICAM) 1.3 m instrument operated by the Small \& Moderate Aperture Research Telescope System (SMARTS) consortium at the Cerro Tololo Inter-American Observatory (CTIO) in Chile. The dual-channel ANDICAM imager utilizes a dichroic filter to separate the beam into its optical and near-IR components, which are then measured by a Fairchild 447 2048 $\times$ 2048 optical CCD and a Rockwell 1024 $\times$ 1024 HgCdTe ``Hawaii" array, respectively. Simultaneity and quality of imaging are achieved as an internal, moveable mirror performs dithering in the near-IR while a single optical exposure integrates \citep{depoy2003}.

We obtained ground-based \emph{VRIJHK} photometry with a cadence of 1--3 d from October 1, 2012 to May 9, 2013 (JD 2456202.8 -- JD 2456422.5). Each optical observation consisted of four 150 s exposures while its IR counterpart was taken in sets of 15 dithered exposures, each 30 s in duration, to avoid saturation. While optical images arrived bias/overscan-subtracted and flat-fielded by SMARTS staff via $\texttt{ccdproc}$, NIR images required further user interaction to flat-field and sky-subtract data using an IRAF script developed by SMARTS personnel, $\texttt{ircombbin}$. We extracted photometry from all local comparison stars and KH 15D via the IRAF task $\texttt{phot}$. Tables \ref{tab:compstarO} and \ref{tab:compstarIR} show standard magnitudes for the optical and near-IR comparison stars, respectively. To maximize signal-to-noise (S/N), we adopted a sky inner radius and width of 13 pix and 5 pix, as well as a 7 pix aperture radius in \emph{VRIK} and 10 pix in \emph{JH}. $K$-band images contained residual flat-fielding errors (``ripples") which set the detection limit at $\sim$ 16 mag. We checked the photometric stability of comparison stars and performed differential photometry using reference stars which exhibited little or no variability. Note that the ANDICAM $I$ filter extends further to the red than  Cousins $I$; accordingly, we performed color transformations on 2010--2013 data using Landolt standards obtained nightly by SMARTS personnel. The 2010--2013 SMARTS $I$ magnitudes were typically 0.1 mag brighter than standard $I$. 

In addition to the 2010--2013 SMARTS observations, we have assembled from the literature for the convenience of future investigators a master table of available standard photometry on KH 15D. Tab.~\ref{tab:masterONIR} shows sample data in the Johnson-Cousins optical \emph{BVRI} \citep{bessell1990} and the near-infrared \emph{JHK} filters. The complete versions are available as ascii downloads online. Note that the optical table contains 2005--2010 SMARTS $I$-band observations that required color transformations to the Cousins system.  We derive from the 2010--2013 SMARTS data a color relation based on the system's $V-I$ value, which attributes a small but significant correction (0.06--0.25 mag) to the SMARTS $I$ magnitude. On average, the 2005--2010 SMARTS $I$ magnitudes were 0.14 mag brighter than standard $I$. In Tab.~\ref{tab:masterONIR} we report $I$ magnitudes on the Cousins system.

\section{Results}
\label{results}
\subsection{2012/13 Light Curves}
\begin{figure}
\epsscale{1.0}
\plotone{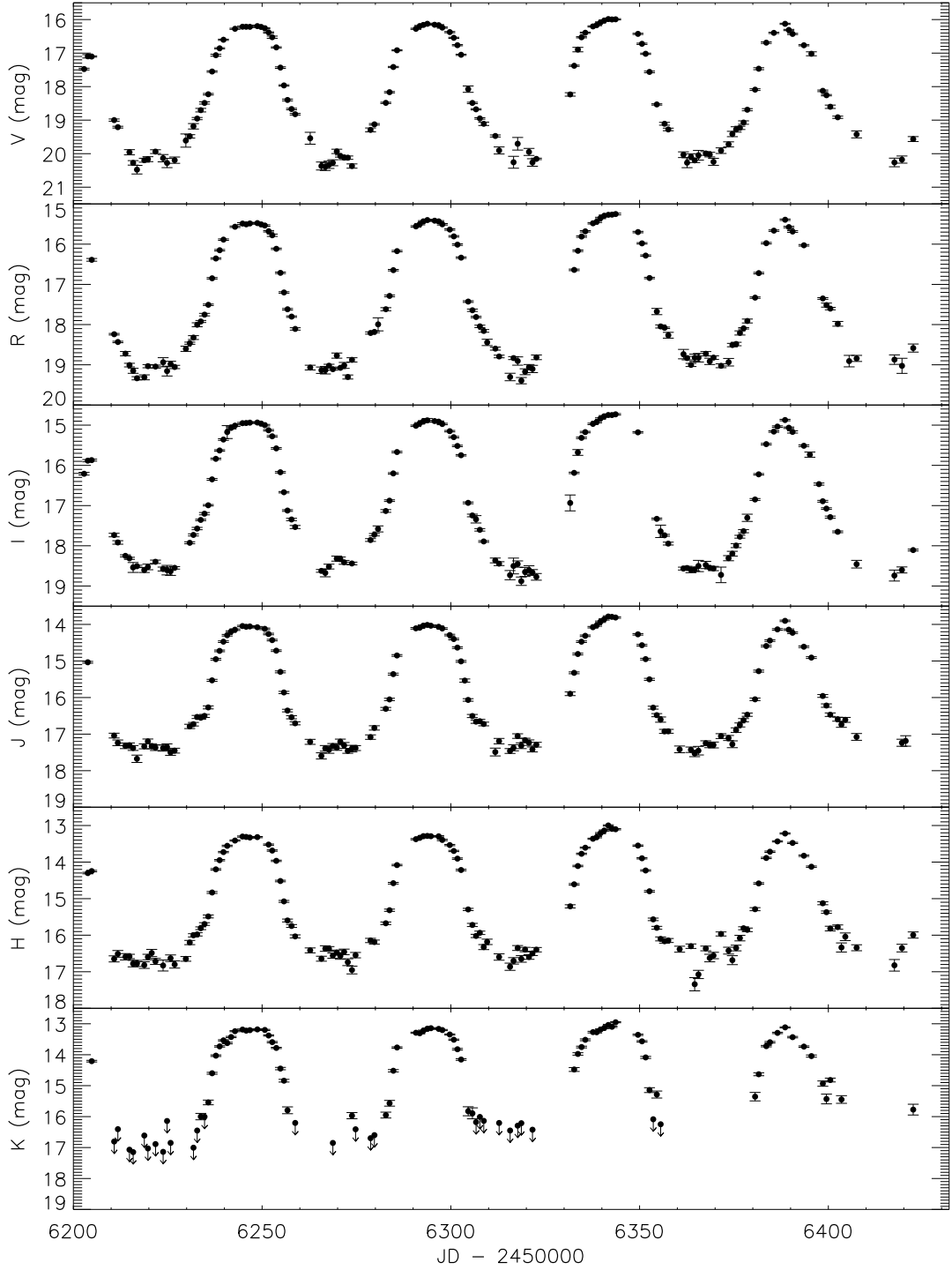}
\caption{The 2012/13 standard \emph{VRIJHK} light curves of KH 15D. Note the anomalous last cycle.}
\label{fig:unfolded}
\end{figure}

This season's unfolded \emph{VRIJHK} light curves (see Fig.~\ref{fig:unfolded}) include over 4 cycles and are characterized by: 1) a maximum which generally becomes progressively more bright and cusped, 2) a sharp drop-off, followed by a shallower, nearly linear decline in brightness during ingress, 3) a short minimum accompanied by a small brightness reversal, and 4) a linear increase, followed by a much steeper rise in brightness during egress. The shape of the last maximum is dramatically different from previous maxima (see Fig.~\ref{fig:unfolded}). The last full cycle contains a well-defined but asymmetric peak that occurs \emph{before} phase = 0.5, followed by a nearly linear decline in magnitude across all bands. We discuss these aberrations further in \S~\ref{discussion}. Except for the anomalous last cycle, the steady increase in the system's maximum brightness is consistent with trends predicted by the precessing ring model. As the screen reveals more of the stellar orbit, more of the stellar photosphere of star B is exposed. 

\begin{figure}
\epsscale{1.0}
\plotone{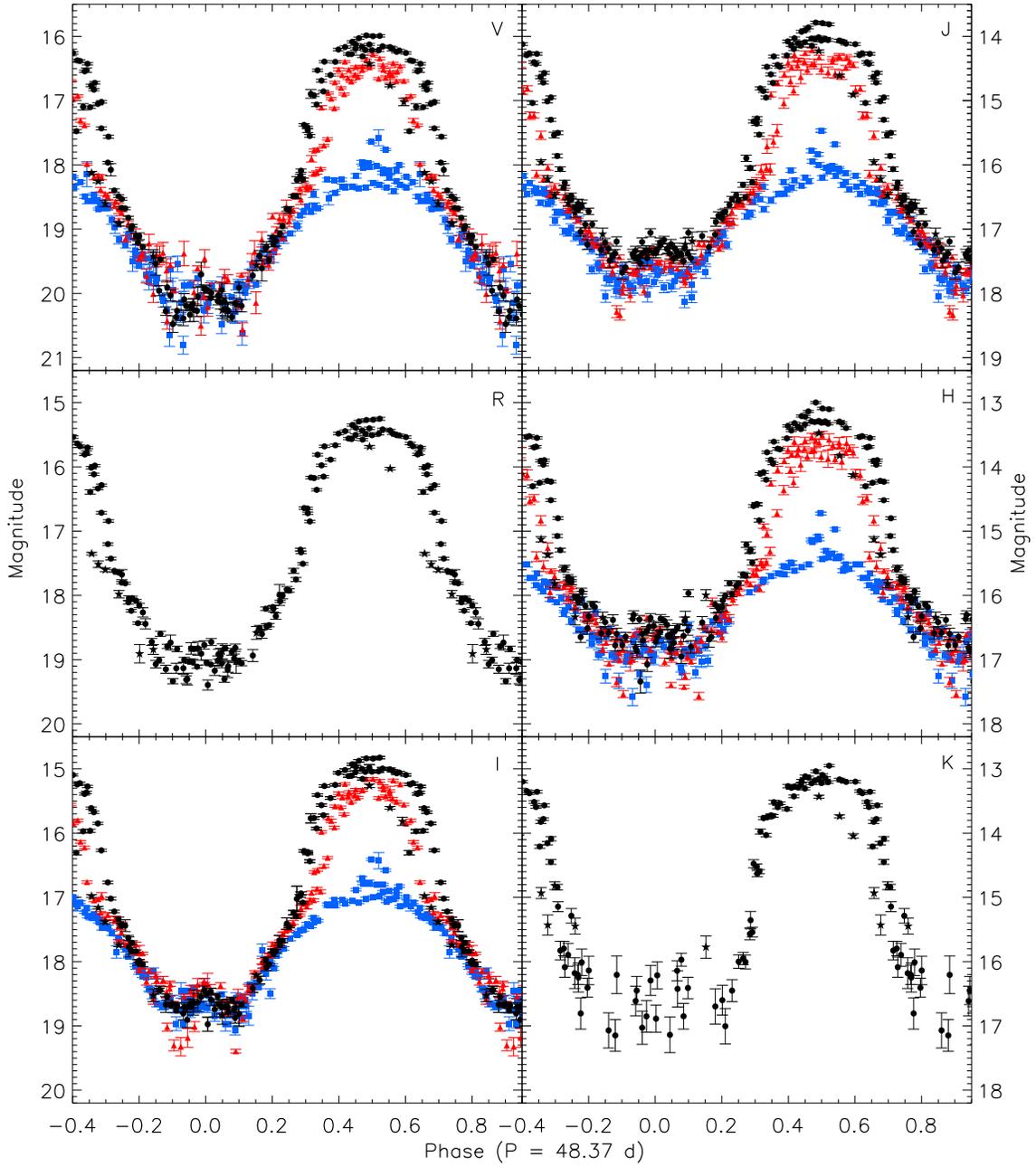}
\caption{\emph{VRIJHK} light curves folded to the binary period of 48.37 d. Blue squares denote data from 2010/11, red triangles are from 2011/12, and black circles indicate 2012/13 data. The anomalous half-cycle discussed in the text is emphasized as black stars.}
\label{fig:phase}
\end{figure}

To better illuminate the periodic features, we phase-fold the light curves to the binary period of 48.37 d in Fig~\ref{fig:phase}. Data from previous SMARTS seasons (2011/2012 = red triangles, 2010/2011 = blue squares) are added for comparison. The small brightness reversals reaching a few tenths of a magnitude are now more readily apparent, especially in $V$, $I$, and $J$; in the other filters, the minima do not have coherent shape but are dominated by large scatter. We note that the peak of central rebrightenings is offset from phase = 0 and that the rise to brightness reversal is shallower than the decline. As described earlier, ingress and egress are characterized by two regimes: steep, more curved profiles closer to maxima and shallow, linear slopes closer to minima. Discontinuities in slope, or points of inflection (POI) occur near $I \approx 17$ mag, where the two regimes intersect (see \S~\ref{discussion}). Since we first began the SMARTS campaign in 2010, the POI has shifted in phase by $\sim$ 0.05 yr$^{-1}$, consistent with the steady advancement of an occulting edge.

\begin{figure}
\epsscale{1.0}
\plotone{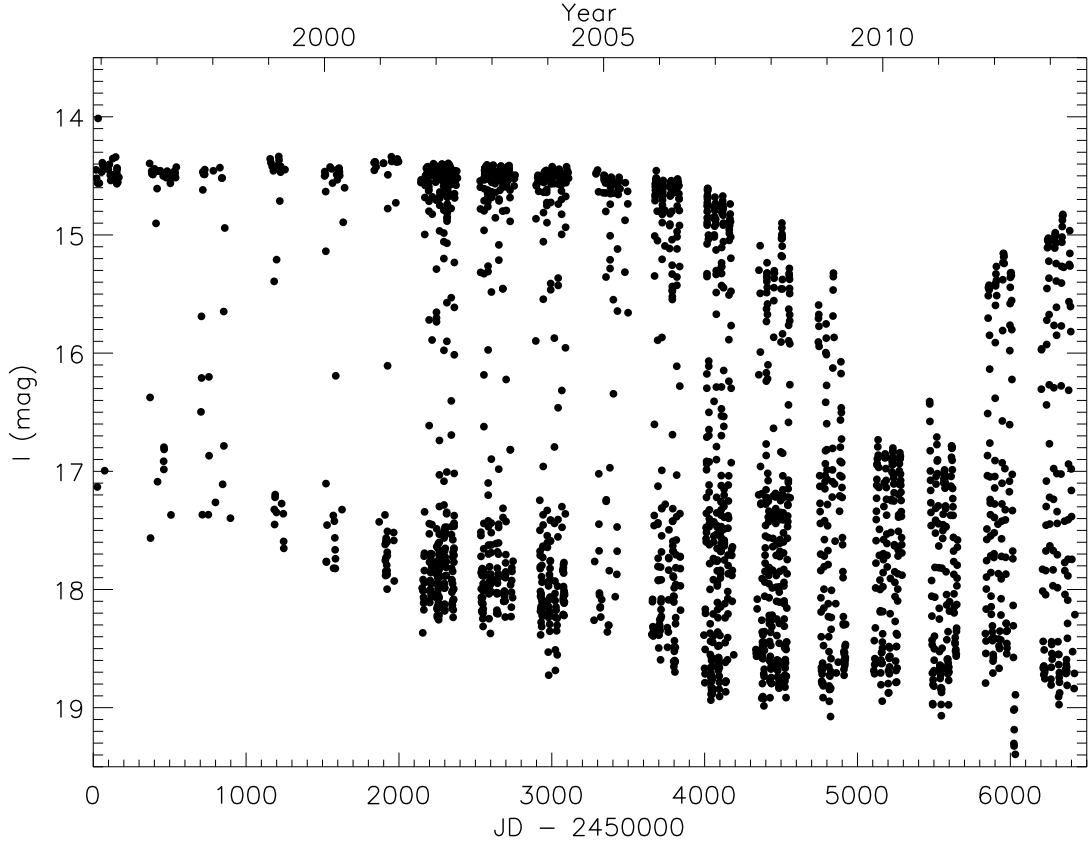}
\caption{The modern light curve of KH 15D. From 1995--2009, star A was fully or partially visible above the occulting edge near apastron. In 2010 and 2011 both stars were fully or nearly fully occulted at all orbital phases. Since 2012, the photosphere of star B is partially seen above the occulting edge during apastron passage. Note that star B was also seen on one night near periastron in 1995. }
\label{fig:longterm}
\end{figure}

We plot the long-term Cousins $I$-band behavior of KH 15D in Fig.~\ref{fig:longterm}. The peak brightness of the system had been steadily decreasing from 2006 (epoch of star A) until it reached a ``faint state" in late 2009 (epoch of complete obscuration). The system then brightened suddenly in late 2011 (epoch of star B), and the steady increase in maxima exhibits some symmetry with the decline in peak brightness seen in earlier data. For the 2012/13 observing season, the system's maximum brightness never exceeded $I = 14.8$ mag, which is $\sim$ 0.3 mag dimmer than the maximum out-of-eclipse flux when the slightly later-type and presumably cooler, fainter K7 star was the source being eclipsed. 

From historical archives we know that the combined light of stars A and B amounted to $I=13.57 \pm 0.03$ mag \citep{winn2006}. Modern (CCD) photometry informs us that star A has an unobscured brightness of $I = 14.47 \pm 0.04$ mag, implying that star B has $I = 14.19 \pm 0.06$ mag. This signifies that star B's photosphere is still significantly obscured by some (projected) portion of the circumbinary ring. Based on the behavior of the upper envelope on the long-term $I$-band light curve, we expect star B to reach $I = 14.4-14.5$ mag next observing season and that it will not rise completely above the ring horizon until around February of 2015.

\subsection{Color Variations}
\begin{figure}
\epsscale{1.0}
\plotone{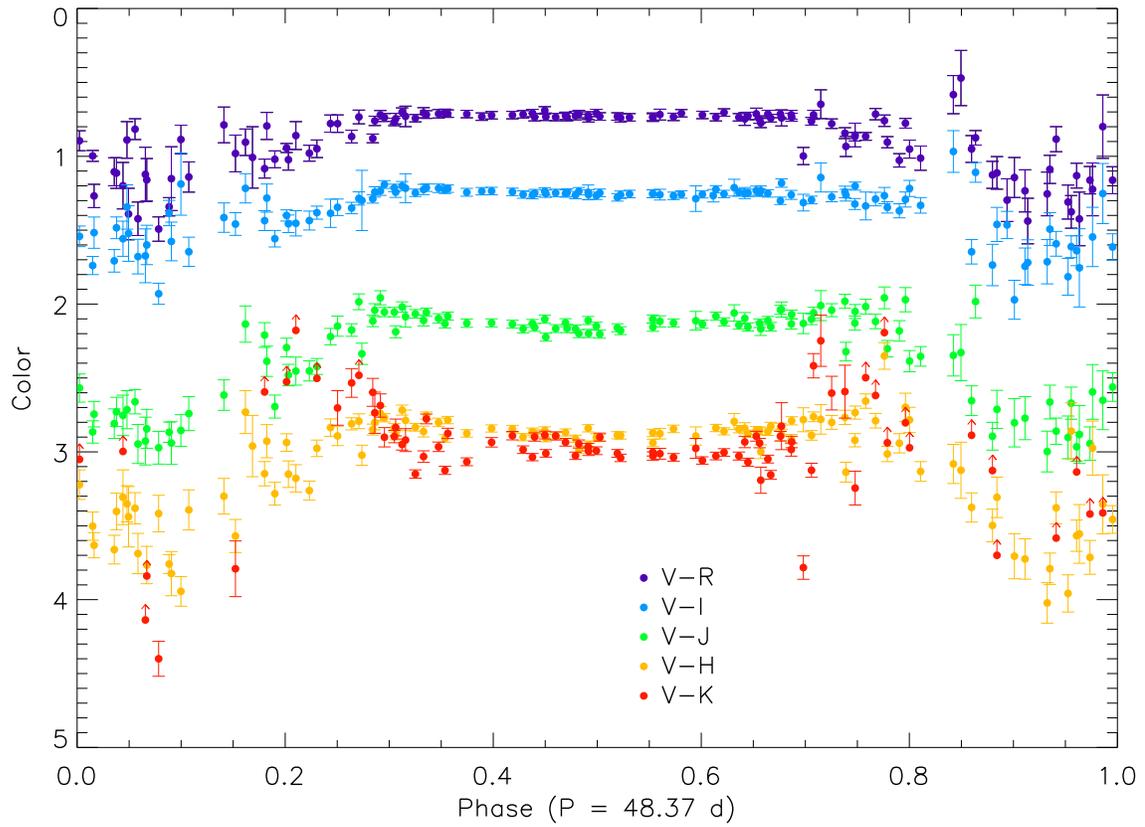}
\caption{2012/13 color vs. phase diagram, based on the ephemeris of \citet{hamilton2005}. Phase 0 is close to periastron and phase 0.5 is close to apastron.}
\label{fig:color}
\end{figure}

Fig.~\ref{fig:color} shows the color variations of KH 15D as a function of phase for the most recent observing season. Notice the flatness of color near maxima (phase = 0.5) when the star is (partially) unobscured near apastron. Blue ``shoulders" are also present at phases 0.3 and 0.7 in the NIR colors, and the amplitude of bluing increases from $V-J$ to $V-K$. Moreover, the system becomes redder by $>$ 0.5 mag in every color near mid-eclipse (phases 0.9--1.0 and 0.0--0.1). The difference in $V-H$ color between in and out of eclipse reaches 1.0 mag. We discuss possible sources for the slight bluing and dramatic reddening in \S~\ref{discussion}. 

Fig.~\ref{fig:longtermcolors} and~\ref{fig:JH} show the system's colors as a function of brightness in magnitudes, where 2003--2005 \emph{JHK} data are from \citet{kusakabe2005} and 2002--2009 \emph{VI} observations are from \citet{hamilton2005} and \citet{herbst2010}. In general, the color-magnitude behavior of KH 15D may be divided into three regimes: 1) bright phase where continuum emission from the stellar photosphere dominates ($I<17$ mag), 2) intermediate phase where the stellar disc is 100\% covered ($17 < I < 18$ mag), and 3) faint phase where the system reddens sharply and ubiquitously ($I>18$ mag). The colors in regime 1 are relatively constant, especially in $V-I$, where the standard deviation in bright phase color is a few hundredths of a magnitude. See Tab.~\ref{tab:syscolor} for a list of bright phase median colors for each epoch where data are available. The color behavior is more complex at intermediate and faint phases. The large scatter in color over the years, despite its chaotic appearance, has a simple explanation, which is presented in \S~\ref{discussion}. In $I-J$ and $I-H$, the colors seem to initially become slightly bluer near $I \approx 17$ mag, consistent with the blue shoulders described earlier. Beyond $I = 17.2$ mag, the colors redden slightly and/or exhibit scatter, until finally, they redden significantly to levels beyond their bright phase values. The degree of reddening at the faintest phases suggests the presence of a cooler third body (see \S~\ref{discussion}). 

\section{Discussion}
\label{discussion}
\subsection{Long-term $V-I$ Evolution}
\begin{figure}
\epsscale{0.9}
\plotone{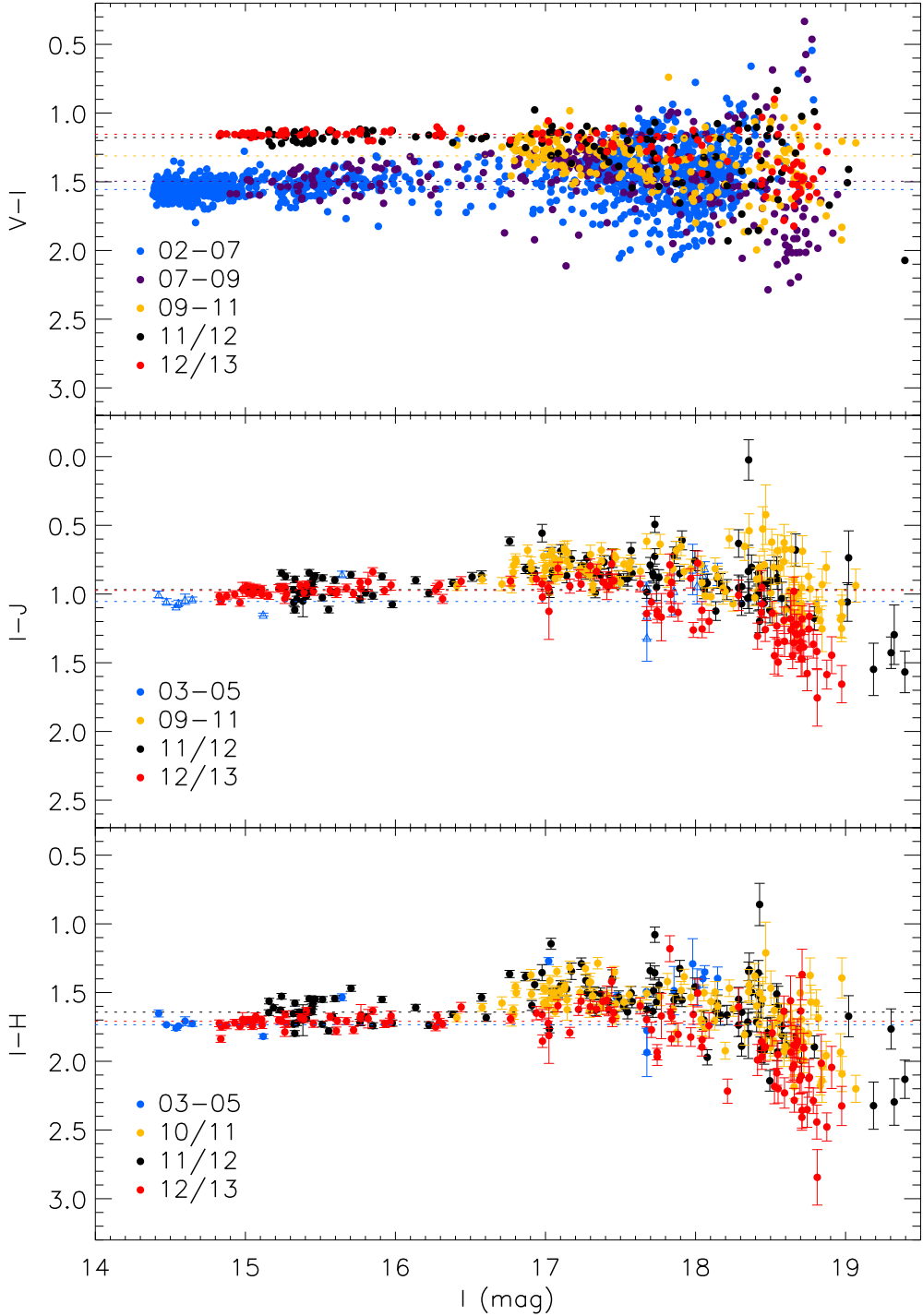}
\caption{Color vs. magnitude ($I$) plots in $V-I$ (top), $I-J$ (middle), and $I-H$ (bottom) during selected observing seasons. The dotted lines indicate the median bright phase colors of the system, where each color corresponds to an observing epoch, as noted in the plot legend. Note the extreme redness in $I-J$ and $I-H$ at faint magnitudes. This is consistent with third light from a putative hot, young Jupiter-mass planet. }
\label{fig:longtermcolors}
\end{figure}

The long-term $V-I$ color-magnitude diagram (see Fig.~\ref{fig:longtermcolors}) clearly demonstrates two distinct bright phase colors corresponding to star A (2002--2009) and star B (2011--2013). This confirms that star B is continuing to rise and set above the occulting horizon. Although star B remains partially covered at maxima, its bright phase $V-I$ is quite stable. We measure a median $V-I$ color of 1.15 $\pm$ 0.02 mag and $1.18 \pm 0.03$ for 2012/13 and 2011/12, respectively. Note that the previously published $V-I$ value of star B (1.28 mag; \citet{capelo2012}) had not been transformed to the standard system. When the color term is taken into consideration, the system's bright phase $V-I$ is closer to the expected color for the spectroscopically identified K1 star. The difference in color between star A (2002--2007) and star B (2012/13) is $\Delta (V-I) = 1.55-1.15 = 0.4$ mag (see Tab.~\ref{tab:syscolor}), which most closely corresponds to the color difference between a K7V and K4V \citep{bessell1990,briceno2007} or that between K7V and K1III \citep{bessell1988}. 

It is evident from the long-term $V-I$ diagram that during recent years (2011--2013), the system at intermediate and faint phases ($I > 17$ mag) tends to become redder than star B's median photospheric color. By contrast, when star A was undergoing regular occultations, the system exhibited significant bluing as it became fainter. The amount of bluing, $\Delta(V-I)$ = $0.3-0.4$ mag, corresponds exactly to the color of star B's photosphere as revealed in recent data. We therefore interpret the bluing apparent in 2002--2009 observations as due to scattered light from (the then fully occulted) star B, the hotter and bluer of the pair. Conversely, we attribute the redness in $V-I$ during 2011--2013 eclipses as due to contaminant light from the now fully occulted star A. The fact that the system color during 2009--2011, when both stars were completely obscured, is bound by the blueness and redness set by star B and star A, respectively, also supports our argument (notice the upper and lower envelopes in Fig.~\ref{fig:longtermcolors}). Indeed, this indicates that forward scattering and selective reddening due to ring particles do not contribute significantly to the large scale $V-I$ behavior. Rather, the observed trends may be understood as a mixture of direct and neutrally scattered light from two stars of somewhat different color. 

It is quite possible that during epochs when star A or star B is partially revealed, the system's \emph{bright} phase color may also be somewhat affected by the other (fully occulted) star's scattered light. Indeed, the data (see Fig.~\ref{fig:longtermcolors} and Tab.~\ref{tab:syscolor}) demonstrate a slightly bluer bright phase color in 2007--2009 when star A was becoming increasingly more covered near apastron than 2002--2007 when star A was fully revealed. If the mixture of direct and scattered light plays a small but significant role even during bright phase, then the 2012/13 bright phase $V-I = 1.15$ mag is only a lower limit on star B's photospheric color. That is, the star would be slightly bluer and would more closely resemble a K1V/K2V star.

\subsection{Cycle to Cycle Variations}
The abrupt linear decline in magnitude from maximum during the last full cycle of this season's observations is intriguing and unprecedented. If the decrease in brightness were due to cold spots on star B, a common feature in weak-lined TTS, then the spot (or collection of spots) would have to cover 50--60\% of the 50\% revealed stellar disc . While this value is consistent with typical covering factors (10--40\%; \citep{herbst1994}), the effect of dark spots on the stellar continuum should decrease with increasing wavelength, given that the temperatures of spots are typically $\sim$ 750 K lower than photospheric values \citep{bouvier1989}. The data, however, show remarkably similar linearity in the steady magnitude decline (6.1 $\pm$ 0.1 mag period$^{-1}$,  $\chi_r^2$ = 1.7) across all filters $V$ through $K$. Although this particular feature is inconsistent with rotational modulation of spots, dark blemishes on star B are not unexpected, given the spottedness of its (presumably coeval) companion \citep{hamilton2005,herbst2010}. 

Perhaps the trailing edge of the screen is not as straight as the leading edge, although the saw-like top of the last cycle indicates an edge just as sharp. Based on the wavelength independence of the early onset of flux decline, a more favorable explanation may be the presence of a local protuberance of material of the occulting edge. This grey attenuation of flux is consistent with the behavior of the system immediately after complete photospheric obscuration, i.e., at the POI. The anomalies and the normal occultations have similar color behavior, suggesting a common cause.

\subsection{Wavelength Dependence of Occultations}

Fig.~\ref{fig:qcurves} shows the system's change in magnitude from median maxima values during 2003--2005 when star A is fully revealed, 2011/12 when star B is $\sim$ 40\% revealed, and 2012/13 when star B is $\sim$ 50\% revealed. The POIs (dotted lines) mark the transition between two distinct regimes of light behavior. We interpret the POIs as second and third contact when star B is at the cusp of 100\% coverage for the following reasons. First, although the POIs steadily shift in phase, they occur at nearly the same brightness level $I\sim17$ mag (see Fig.~\ref{fig:phase} and Fig. 4 of \citet{herbst2010}; note that for the latter plot, 2005--2010 SMARTS data were not transformed to Cousins $I$, and so a 0.1--0.2 mag offset may be expected). Second, the median maximum brightness between 2009--2011, when scattered light from the fully obscured binary dominated the system, is consistent with $I=17$ mag. Finally, at light levels dimmer than $I=17$ mag, system colors deviate significantly from photospheric values, which suggest intermediate and faint phase light is different from bright phase light (see Fig.~\ref{fig:longtermcolors}). 

Interestingly, the depth of eclipse from maximum to POI is approximately wavelength independent in the optical and near-infrared (see Fig.~\ref{fig:qcurves}), i.e., light attenuation is grey. While the larger $\Delta$mag value in $K$ in 2012/13 is enticing, it is not a significant detection since the POI is near the S/N limit at $K\sim16$ mag. Neither grey scattering nor grey transmission is compatible with the behavior of interstellar dust, which is typically sub-micron in size. That the attenuation of light for both star A and star B is grey out to $\sim K$ indicates the line of sight material is larger than the size of the wavelength of observation, i.e., grain radii $> 1\ \mu$m \citep{draine2011}. Additionally, the particles in the occulting ring must be similar in size at least on angular scales of the projected binary orbit (which is physically $\sim0.5$ AU across), since star A probes a different line of sight from that of star B.

The color data indicate that when viewing the limb of star B at egress or ingress, the system appears quite blue in many of the colors, especially in $V-H$ and $V-K$ (see Fig.~\ref{fig:color}). The effect may be somewhat asymmetric, appearing stronger during egress. The bluing trend is opposite to what one would expect from limb darkening in an ordinary K star atmosphere. While we have no reason to doubt the integrity of the data at these phases, we also have no simple explanation for them. Perhaps the effects of external heating on star B's atmosphere (from accretion, chromospheric and coronal radiation, and star A) lead to a temperature inversion that manifests itself at short wavelengths as limb brightening. Perhaps forward scattered light plays a role. Or, perhaps the spotted surface of star B is such that we just happen to be seeing cooler (spot) material towards the center and hotter material towards the limb. Additional seasons of monitoring during which a full range of colors is observed should shed light on which of these explanations, if any, is correct.

\subsection{Third Body?}

\begin{figure}
\epsscale{1.0}
\plotone{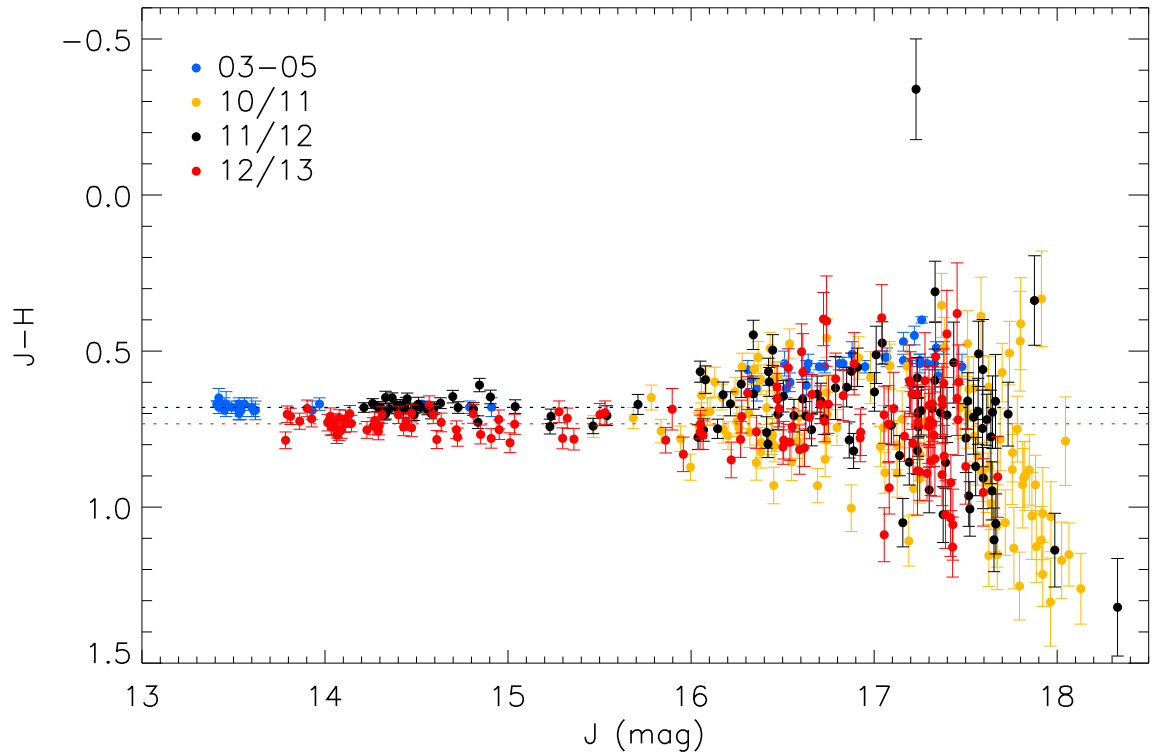}
\caption{$J-H$ vs. $J$. Notice the similarity in median bright phase colors between 2003--2005 (star A 100\% revealed) and 2011/12 (star B 40\% revealed), as well as the slight \emph{reddening} in 2012/13 (star B 50\% revealed) bright phase data.}
\label{fig:JH}
\end{figure}

\begin{figure}
\epsscale{0.8}
\plotone{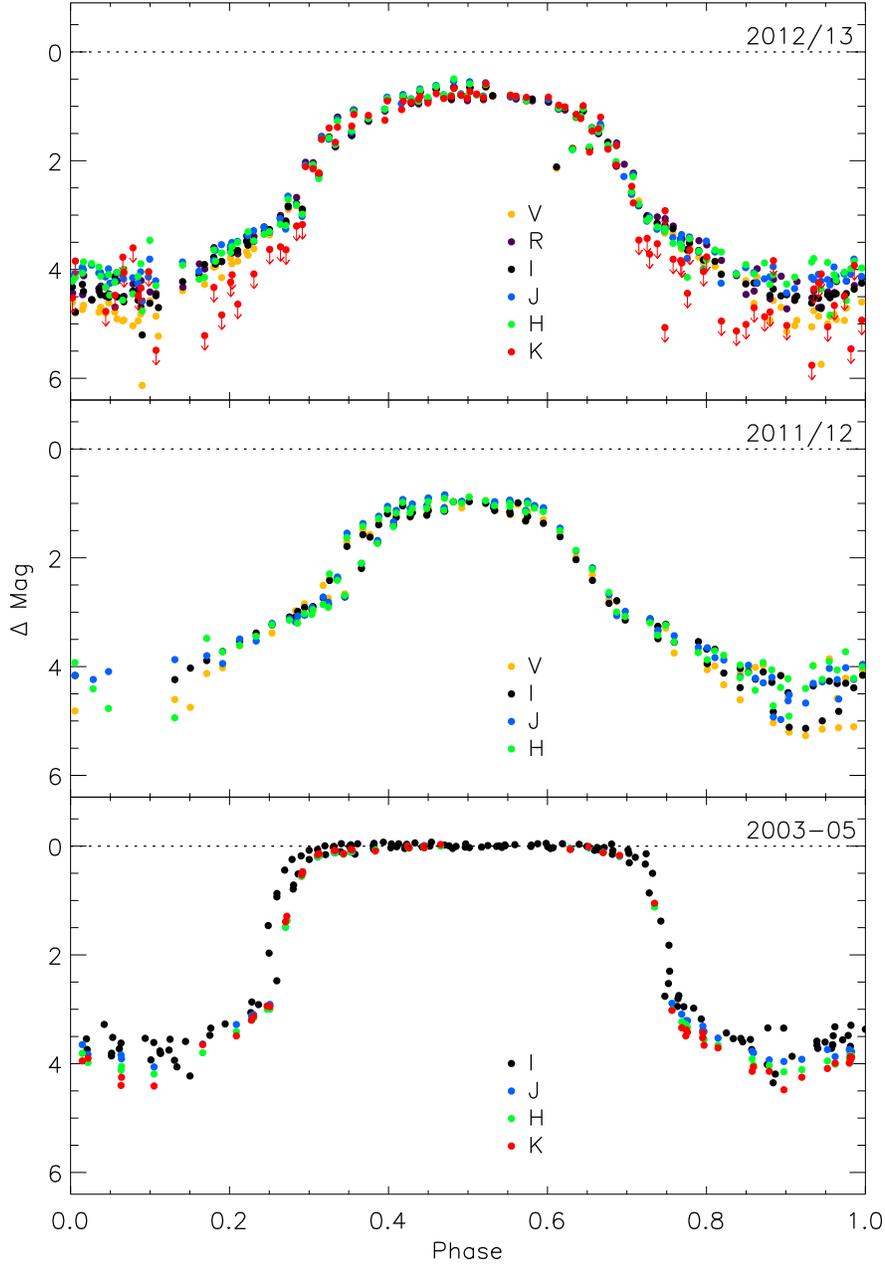}
\caption{The change in magnitude from maximum light (i.e., 0\% photospheric coverage) for KH 15D at various optical and near-infrared wavelengths. Top panel shows \emph{VRIJHK} data from 2012/13, where star B goes from 50\% coverage at maxima to 100\% coverage at the point of inflection (POI; depth of eclipse = 2.2 mag).  Middle panel contains \emph{VIJH} data from 2011/12, where star B goes from 60\% coverage at maxima to 100\% coverage at POI (depth of eclipse = 2 mag). Bottom panel shows \emph{IJHK} data from 2003--2005, when star A rose and set completely (depth of eclipse = 3 mag). Dashed lines represent the maximum light of star B ($I$ = 14.2 mag; top and middle panels) and of star A($I$ = 14.5 mag; bottom panel). Notice that although the depth and duration of eclipse at the POI has changed from year to year, there is little, if any, wavelength dependence within each observing season. }
\label{fig:qcurves}
\end{figure}

The long-term O/NIR color plots (see Fig.~\ref{fig:longtermcolors}) compare the color behavior between the epochs of star A and star B. It is apparent that the colors redden dramatically at the faintest phases (beyond $I\sim18.3$ mag), especially during the most recent season (2012/13) in $I-J$ and $I-H$. We have established earlier that the reddening seen in 2011--2013 $V-I$ at faint phases is likely due to the redness of the now invisible star A, complemented by the bluing seen in 2002--2009, when scattered light from the bluer but fully obscured star B dominated the system at faint phases. Here, we refer to the severe reddening -- beyond the photospheric values of either star -- seen at the faintest phases. At $I\sim18.8$ mag, $I-J$ and $I-H$ colors reach +1.5 and +2.3 mag, respectively, which correspond to color excesses of E($I-J$) = 0.6 mag and E($I-H$) = 0.7 mag. These reddening slopes for KH 15D are much steeper than one would expect for interstellar reddening -- more than twice the values \citep{draine2011}. We also observe that the 2012/13 data show a marginally \emph{redder} (0.05 $\pm$ 0.03) bright phase $J-H$ color (see Fig.~\ref{fig:JH}) than both 2011/12 and 2003--05 data, when star B was $\sim20\%$ more covered and the slightly later-type star A was being eclipsed, respectively. The simplest explanation may be that we are detecting excess light, especially in $J$ and $H$, when the overall system is very faint. 

The existence of a shepherding body massive enough to constrain, but not disrupt, the ring (such as a young gas giant) is well motivated. Physical models of KH 15D necessitate the presence of a truncation mechanism at $\sim$ 5--10 AU to maintain the finite width, warp, and precession of the circumbinary ring. A young super-Jupiter orbiting in (or near) the plane of the central binary should leave observable signatures such as excess near-infrared radiation possibly regulated by the orbital period of the planet. 

According to ``hot start" models, a $\sim$ 1 Myr old 10 M$_J$ planet would have an absolute $M_H=+8$ mag \citep{fortney2008,burrows2012}. Given the distance to NGC 2264, such an object will have $H\sim17$ mag and inject an amount of flux consistent with the potential $H$ excess observed during bright phase. Thermal radiation from a young super-Jupiter is furthermore consistent with the observed degree of reddening at the faintest phases. Finally, the 2003--2005 data show little or no hint of reddening. While this is due to incomplete $I$-band coverage in Fig.~\ref{fig:longtermcolors}, the $J-H$ vs. $J$ plot (see Fig.~\ref{fig:JH}) clearly demonstrates that, despite similar \emph{JH} ranges, the system does not redden beyond the bright phase colors of star A in 2003--2005. This phenomenon may be explained by the fact that if there were a third body, then every few years or so the planet would be on the far side of the ring and its radiation signatures would be completely obscured. We emphasize, however, that to test this ring confinement theory requires spectroscopic, i.e., radial velocity, confirmation.

\subsection{Ring Confinement via Clumping Instability?}

In recent years, a new mechanism has been proposed for ring formation in dusty disks with sufficient gas \citep{klahr2005,besla2007,lyra2013}, without requiring shepherding planets. Our data does not preclude this photoelectric instability as a a possibility for ring confinement. This formation scenario occurs when dust heats gas in the circumstellar or circumbinary disk via photoelectric heating, creating local pressure maxima which concentrate the solid particles.  \citet{lyra2013} modeled this clumping instability and successfully produced narrow, eccentric ring structures similar to those observed in debris disks around Formalhaut and HD 61005. Although a gas-to-dust ratio of order unity is enough for this confinement mechanism to occur, the gas-to-dust ratio in the KH 15D disk is currently not well constrained \citep{lawler2010}, and thus we do not know whether there is sufficient gas for this instability. While the near-IR excess we find during minimum light might be explained by radiation from grains composing the ring, the inferred temperatures of 1000--2000 K seem too high for a ring located 1--5 AU from the central stars. Near-infrared spectral data and longer wavelength flux measurements would be useful in evaluating this alternative model.

\section{Conclusion}
\label{conclusion}
The serendipitous alignment of KH 15D allows a unique, but time-limited opportunity to probe the properties of this T Tauri system and its circumbinary ring. The most recent O/NIR data reveal the continued ``secular" rise of star B's photosphere above the circumbinary ring's obscuring horizon. It has moved from $\sim40\%$ revealed during 2011/12 to $\sim50\%$ revealed during 2012/13. The multi-wavelength data set demonstrates regular and dramatic reddening in $I-H$, $I-J$, and $J-H$ at faint phases, possibly caused by a luminous giant planet. We show that the flux attenuation at second contact is grey from $V$ to $K$, which indicates the line of sight particles are larger than $\sim 1\ \mu$m, i.e., significantly larger than interstellar dust. Finally, we have discovered an apparent protuberance in the trailing edge of the occulting screen, which significantly impacted the shape of the light curve during the last full cycle of the 2012/13 observing season.

We emphasize that our current data set does not demonstrate the detection of a young planet; it merely hints at one. We urge continued photometric and spectroscopic monitoring of the system in the near-infrared to help disambiguate the features reported here. The detection of a gas giant in such a young binary star system would provide critical observational guidance to models of planet formation.

\acknowledgments

We wish to acknowledge our collaborators C. Hamilton and H. Capelo for many helpful discussions relevant to this work. We thank the observers and personnel at the SMARTS consortium. This work was partially funded by a NASA grant to W. Herbst through the Origins of Solar Systems program, and partially supported by grants through Wesleyan University.

\clearpage

\begin{deluxetable}{cccccccc}
\tabletypesize{\scriptsize}
\tablewidth{0pt}
\tablecaption{\emph{VRI} Magnitudes for Optical Comparison Stars}
\small
\tablehead{
\colhead{Star (\citet{hamilton2005} ID)} & \colhead{V} & \colhead{$\Delta$V} & \colhead{R} & \colhead{$\Delta$R} & \colhead{I} & \colhead{$\Delta$I} }

\startdata
1 (F) & 13.869 & 0.020 & 13.375 & 0.021 & 12.902 & 0.021 \\
2 (C) & 12.969 & 0.020 & 12.617 & 0.020 & 12.240 & 0.021 \\
3 (D)\tablenotemark{\dagger} & 15.013 & 0.020 & 14.328 & 0.021 & 13.645 & 0.021 \\
\enddata

\tablenotetext{\dagger}{Star 3 exhibited significant flux variability this year and was therefore not used for calibration.}
\label{tab:compstarO}
\end{deluxetable}

\clearpage

\begin{deluxetable}{ccccccccc}
\tabletypesize{\scriptsize}
\tablewidth{0pt}
\tablecaption{\emph{JHK} Magnitudes for Infrared Comparison Stars}
\small
\tablehead{
\colhead{Star} & \colhead{J} & \colhead{$\Delta$J} & \colhead{H} & \colhead{$\Delta$H} & \colhead{K} & \colhead{$\Delta$K} & \colhead{SIMBAD Identifier} & \colhead{\tablenotemark{\dagger}} }

\startdata
 01 & 15.304 & 0.048 & 14.564 & 0.045 & 14.285 & 0.073 & Cl* NGC 2264 FMS 2-1314 & \emph{JH} \\ 
 02 & 13.422 & 0.030 & 12.725 & 0.023 & 12.549 & 0.026 & V* OS Mon & \emph{JHK} \\ 
 03 & 12.205 & 0.024 & 11.541 & 0.024 & 11.356 & 0.023 & V* V427 Mon & \emph{JHK} \\ 
 04 & 13.110 & 0.023 & 12.336 & 0.023 & 12.181 & 0.024 & V* V816 Mon & \emph{JHK} \\ 
 05 & 13.415 & 0.024 & 12.548 & 0.023 & 11.928 & 0.021 & V* OQ Mon & -- \\ 
 06 & 16.018 & 0.109 & 14.740 & 0.070 & 13.656 & 0.079 & Cl* NGC 2264 FMS 2-1323 & -- \\ 
 07 & 14.877 & 0.060 & 14.254 & 0.042 & 13.839 & 0.100 & Cl* NGC 2264 FMS 2-1347 & \emph{JH} \\
\enddata
\tablenotetext{\dagger}{Denotes the comparison star's photometric stability in a particular filter and consequently its usefulness as a standard magnitude calibrator.}

\label{tab:compstarIR}
\end{deluxetable}

\clearpage

\begin{center}
\begin{deluxetable}{ccccccccccccccccc}
\rotate
\tabletypesize{\scriptsize}
\singlespace
\tablewidth{0pt}
\tablecaption{Standard \emph{BVRIJHK} Photometry of KH 15D}
\tablehead{\colhead{JD-2400000} & \colhead{$B$} &\colhead{$\sigma$} & \colhead{$V$} &\colhead{$\sigma$} & \colhead{$R$} &\colhead{$\sigma$} & \colhead{$I$} &\colhead{$\sigma$} & \colhead{$J$} &\colhead{$\sigma$} & \colhead{$H$} &\colhead{$\sigma$} & \colhead{$K$} &\colhead{$\sigma$} & \colhead{Obs.} &\colhead{Pap.} 
}
\startdata
33658.9&99.999&9.999&99.999&9.999&99.999&9.999&  13.4&  0.4&99.999&9.999&99.999&9.999&99.999&9.999&     POSS&    J05 \\
52615.77&17.438&0.013&16.050&0.008&15.258&0.008&14.476&0.006&99.999&9.999&99.999&9.999&99.999&9.999&     USNO&    H05 \\
53384.34&99.999&9.999&99.999&9.999&99.999&9.999&14.598&0.032&13.560&0.020&12.890&0.000&12.610&0.010&WISE+IRSF&H10+K05 \\
54153.60&99.999&9.999&18.930&0.102&99.999&9.999&17.859&0.049&99.999&9.999&99.999&9.999&99.999&9.999&     CTIO&    H10 \\
56422.48&99.999&9.999&19.565&0.073&18.584&0.100&18.213&0.024&99.999&9.999&15.996&0.085&15.774&0.174&     CTIO&    W13 \\

\enddata
\tablecomments{Missing data points and uncertainties are signified by 99.999 and 9.999, respectively. Column 16 denotes the observatory from which the photometry was obtained. Column 17 shows the paper from which data was presented (J04 = \citet{johnsonwinn2004}, J05 = \citet{johnsonwinn2005}, H05 = \citet{hamilton2005}, K05 = \citet{kusakabe2005}, H10 = \citet{herbst2010}, W13 = this paper). Entries in columns 16 and 17 which contain a '+' symbolizes different sources for optical and near-infrared data.}

\label{tab:masterONIR}
\end{deluxetable}
\end{center}

\clearpage

\begin{deluxetable}{cccc}
\tabletypesize{\scriptsize}
\singlespace
\tablewidth{0pt}
\tablecaption{Long-term System Colors}
\tablehead{\colhead{Color} & \colhead{Observations} &\colhead{Bright Phase} \\
\colhead{(mag)} & \colhead{(year)} &\colhead{Median (mag)} }
\startdata
$V-I$ & 2002--07 & $1.55 \pm 0.13$ \\
$V-I$ & 2007--09 & $1.48 \pm 0.09$ \\
$V-I$ & 2009--11 & $1.31 \pm 0.18$ \\
$V-I$ & 2011/12 & $1.18 \pm 0.03$  \\
$V-I$ & 2012/13 & $1.15 \pm 0.02$  \\
$I-J$ & 2003--05 & $1.05 \pm 0.09$ \\
$I-J$ & 2011/12 & $0.97 \pm 0.08$  \\
$I-J$ & 2012/13 & $0.97 \pm 0.04$  \\
$I-H$ & 2003--05 & $1.73 \pm 0.08$  \\
$I-H$ & 2011/12 & $1.64 \pm 0.08$  \\
$I-H$ & 2012/13 & $1.71 \pm 0.04$  \\
$J-H$ & 2003--05 & $0.68 \pm 0.01$  \\
$J-H$ & 2011/12 & $0.68 \pm 0.02$  \\
$J-H$ & 2012/13 & $0.73 \pm 0.03$  \\

\enddata
\label{tab:syscolor}
\end{deluxetable}

\clearpage

\end{document}